\documentclass[a4paper,usenatbib]{mnras}
\usepackage{newtxtext,newtxmath}
\usepackage[T1]{fontenc}
\usepackage{ae,aecompl}

\usepackage{graphicx}	% Including figure files
\usepackage{amsmath}	% Advanced maths commands
\usepackage{amssymb}	% Extra maths symbols
\usepackage{epsfig,color}
\usepackage[all]{xy}
\usepackage{mathrsfs}
\usepackage{ulem}
\usepackage{empheq}
\usepackage{bm}
\usepackage{harpoon}
\usepackage{multirow}
\usepackage{deluxetable}

\def\calA{{\cal A}}
\def\ergs{\rm erg\,s^{-1}}
\def\kms{\rm km\,s^{-1}}
\def\sunm{M_{\odot}}

\def\bhm{M_{\bullet}}
\def\calS{{\cal S}}
\def\fBLR{f_{\rm _{BLR}}}
\def\VFWHM{V_{_{\rm FWHM}}}

\title[Measuring black hole mass]{Measuring black hole mass of type I active galactic nuclei by spectropolarimetry}

\author[Yu-Yang Songsheng \& Jian-Min Wang]{
Yu-Yang Songsheng$^{1,\,3}$ and
Jian-Min Wang$^{1,\,2,\,3}$\thanks{E-mail: wangjm@ihep.ac.cn}
\\
$^{1}$Key Laboratory for Particle Astrophysics, Institute of High Energy Physics, Chinese Academy of Sciences, 
19B Yuquan Road, \\ Beijing 100049, China\\
$^{2}$University of Chinese Academy of Sciences, 19A Yuquan Road, Beijing 100049, China\\
$^{3}$National Astronomical Observatories of China, Chinese Academy of Sciences, 20A Datun Road, Beijing 100020, China
}

\pubyear{2017}

\begin{document}
\label{firstpage}
\pagerange{\pageref{firstpage}--\pageref{lastpage}}
\maketitle

% Abstract of the paper
\begin{abstract}
Black hole (BH) mass of Type I active galactic nuclei (AGN) can be measured or estimated through 
either reverberation mapping (RM) or empirical $R-L$ relation, however, both of them suffer from 
uncertainties of the virial factor ($\fBLR$), thus limiting the measurement accuracy. In 
this letter, we make an effort to investigate $\fBLR$ through polarised spectra of the 
broad-line regions (BLR) arisen from electrons in the equatorial plane. Given the BLR
composed of discrete clouds with Keplerian velocity around the central BH, we simulate 
a large number of  spectra of total and polarised flux with wide ranges of parameters of 
the BLR model and equatorial scatters. We find that the $\fBLR$-distribution of polarised 
spectra is much narrower than that of total ones. This provides a way of n accurately estimating 
BH mass from single spectropolarimetric observations of type I AGN whose equatorial
scatters are identified.
\end{abstract}


%%%%%%%%%%%%%%%%%%%%%%%%%%%%%%%%%%%%%%%%%%%%%%%%%%

%%%%%%%%%%%%%%%%% BODY OF PAPER %%%%%%%%%%%%%%%%%%

\section{Introduction}

Reverberation mapping (RM) is nowadays the most common technique of measuring black hole (BH) mass of type I 
active galactic nuclei (AGNs) except for a few local AGNs spatially resolved \citep{peterson1993reverberation, peterson2014measuring}. 
RM measures time lags $\Delta t$ of broad emission line with respect to varying continuum as ionizing photons,
allowing us to obtain the emissivity-averaged distance from the BLR to the central black hole. Assuming fully 
random orbits of the BLR clouds with Keplerian velocity, we have the BH mass as 
\begin{equation}
\bhm  = \fBLR \frac{c\Delta t\VFWHM^2}{G},
\end{equation}
where $\fBLR$ is the virial factor,
$\VFWHM$ is the full-width-half-maximum of the broad emission line profiles, $c$ is the light speed and $G$ is the
gravity constant. The total error budget on the BH mass can be simply estimated by 
$\delta \bhm/\bhm \approx \left[\left(\delta \ln \fBLR\right)^2+0.08\right]^{1/2}$,
where $\Delta t$ and $\VFWHM$ are usually of 20\% and 10\% for a typical measurement of RM observations,
respectively. Obviously, the major uncertainty on the BH mass is due to $\fBLR$, however, it could be different by 
a factor of more than one order of magnitude \citep{pancoast2014modelling2}. 
Its dependence on kinematics, geometry and inclination of the BLR is poorly understood
\citep{krolik2001systematic,collin2006systematic}.

For those AGNs with measured stellar velocity dispersions $\sigma_{*}$ of bulges and RM data, 
$\fBLR$ can be calibrated by the $\bhm - \sigma_{*}$ relation found in inactive 
galaxies\citep{onken2004supermassive, woo2010lick}.
$\langle \fBLR \rangle$ as an averaged one can only remove the systematic bias between the virial 
product $c\Delta tV_{\rm _{FWHM}}^2 / G$ and $M_{\bullet}$ for a large sample, however, $\fBLR$ is 
poorly understood individually. Furthermore, the zero point and scatters of the 
$M_{\bullet} - \sigma_{*}$ relation depend on bulge types of the host galaxy, and virial factors 
of classical bulges and pseudobulges can differ by a factor of $\sim 2$ \citep{ho2014black}. The 
calibrated values of $\fBLR$ lead to $\delta\bhm/\bhm\sim 2$, yielding only rough estimations of 
BH mass in AGNs.

Recently, a motivated idea to test the validity of $\fBLR$ factor has been suggested 
by \citet{du2017hidden} in type II AGNs through the polarised spectra. In principle, the polarised
spectra are viewed with highly face-on orientation to observers and $\fBLR$ in type II AGNs should 
be the same as with type I AGNs. They reach a conclusion of $\fBLR \sim 1$ from a limited sample. 
For type I AGNs, the polarised spectra received by a remote observer correspond to ones viewed by 
an edge-on observer, lending an opportunity to measure BH mass similar to cases of NGC 4258 through 
water maser \citep{miyoshi1995evidence} or others through CO line \citep{barth2016measurement}.

\begin{figure*}
\centering
\includegraphics[scale=0.45]{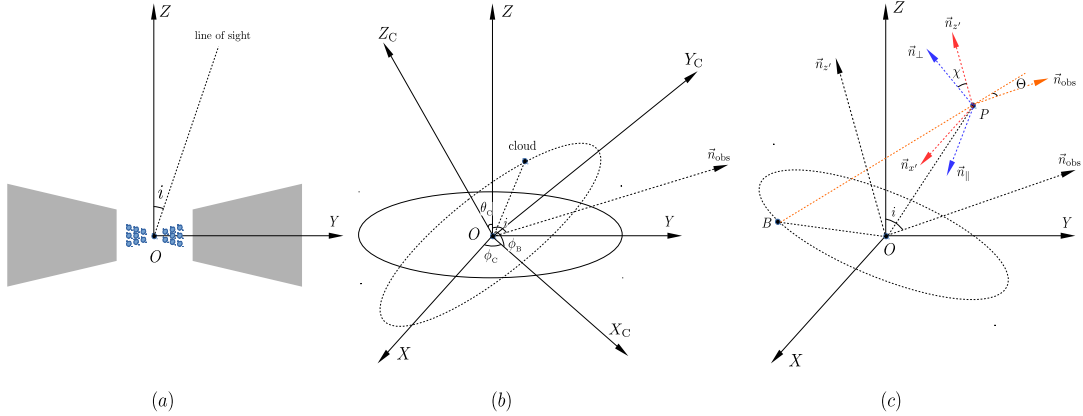}
\caption {
Panel {\it a} is a cartoon of a type I AGN with an equatorial scattering region. 
The blue points represent clouds in the BLR and the grey region the scatters
on the equatorial plan. $i$ is the inclination angle to a remote observer in the $O-YZ$-plane. 
Panel {\it b} is the frame for the BLR geometry. $O-XY$ is the equatorial plane. 
$O-X_{\rm _{C}}Y_{\rm _{C}}$ is the orbital plane of one specific cloud, which can 
be obtained by rotating $X$ around $Z$ by $\phi_{\rm _{C}}$ and then rotating $Z$ 
around $X_{\rm _{C}}$ by $\theta_{\rm _{C}}$. 
The phase angle of the cloud relative to $OX_{\rm _{C}}$ is $\phi_{\rm _{B}}$. 
Panel {\it c} is the scattering geometry used here. ${\bf BP}$ is the incident light from 
point $B$ on one orbit. $\vec{n}_{\rm obs}$ is the direction of sight of 
the observer (i.e. the direction of scattered light). The vectors of
$\vec{n}_{\rm obs}$, $\vec{n}_{\bot}$, $\vec{n}_{\parallel}$, $\vec{n}_{z'}$ and 
$\vec{n}_{x'}$ are explained in Appendix.}
\end{figure*}
\vglue -0.2cm

In this letter, we investigate $\fBLR$ in type I AGNs through modelling the scattering 
polarised spectra quantitatively. In section 2, we build a dynamical model for BLR and scattering region 
of Type I AGNs for polarized spectra. In section 3, we simulate a large number of spectra for a large
range of model parameters to get the distribution of $\fBLR$ for both total and polarized spectra. We
find that $\fBLR$ is in a very narrow range for polarised spectra.
In section 4, we draw conclusions and discuss potential ways of improving the accuracy of BH mass determination.

\section{Polarized spectra from of equatorial scatters}
Optical spectropolarimetric observations of type II AGNs discover that there is a broad component 
of emission line in polarised spectra, indicating appearance of: 1) a BLR hidden by torus; 2) at 
least one scattering region outside the 
torus \citep{antonucci1985spectropolarimetry, tran1992detection, miller1991multidirectional}.
Radio observations also show that radio axes of most type II AGNs are nearly perpendicular to the 
position angle of polarization \citep{antonucci1983optical, brindle1990optical}, showing that 
scattering regions of Type II AGNs situated outside the torus but aligned with the axes of the 
AGNs, called polar scattering region. In contrast, observations of type I AGNs reveal that 
position angles of the polarization are more often aligned with radio 
axes \citep{antonucci1983optical, antonucci1984optical,smith2002spectropolarimetric}. 
Equatorial scattering regions may exist, which are hidden by the torus, 
but can be seen in the polarised spectra of type I AGNs \citep{smith2005equatorial}.

We follow the geometry of equatorial scatters as in \citet{smith2005equatorial}. The geometry 
of the scattering regions and BLR are shown in Fig. 1{\it a}. The details of the geometric 
relations are provided in the Appendix. If the half opening angle of the scattering region is 
$\Theta_{\rm _{P}}$, the inner and outer radius of the scattering region is 
$r_{\rm _{P,i}} \mbox{and } r_{\rm _{P,o}}$, then we have 
$r_{\rm _{P}}\in[r_{\rm _{P,i}},r_{\rm _{P,o}}],\theta_{\rm _{P}}\in[\pi/2-\Theta_{\rm _{P}},\pi/2+\Theta_{\rm _{P}}]$. 
Scatterings caused by inter-cloud electrons in the BLR have been estimated as
$\tau_{_{\rm BLR}}\approx 0.04\,R_{\rm 0.1pc}$ [see Eq. (5.13) in \citet{krolik1981two}], 
where $R_{\rm 0.1 pc}=R_{\rm BLR}/0.1\,{\rm pc}$ is a typical size of the BLR (Bentz et al. 
2013), which can be totally neglected for polarised spectra. We thus assume that the scattering 
region is composed of free cold electrons beyond the BLR, and the whole region is optically thin, 
i.e. the optical depth $\tau_{\rm es} \sim \bar{n}\ell\sigma_{\rm T} < 1$, where $\bar{n}$ is the 
average number density of the electrons, $\ell$ is the typical scale of the region and 
$\sigma_{\rm T}$ is the Thomson cross section. The distribution of the number
density of electrons is assumed to be a power law as
$n_{\rm _{P}}(r_{\rm _{P}}, \theta_{\rm _{P}},\phi_{\rm _{P}}) = n_{\rm _{P0}} \left(r_{\rm _{P}}/r_{\rm _{P,i}}\right)^{-\alpha}$,
where $n_{\rm _{P0}}$ is the number density at inner radius $r_{\rm _{P,i}}$.

We assume BLR is composed of a large quantity of independent clouds rotating around the black hole, and has 
a geometry indicated by Fig. 1{\it b} \citep{pancoast2011geometric, li2013bayesian, pancoast2014modelling1}.
The detailed geometric relations of BLR are given in Appendix. 
Suppose that the unit of length is $R_{\rm g} \equiv GM_{\bullet}/c^2$ and orbits of the clouds are circular. 
The velocity of the cloud is
\begin{equation}
\vec{v}_{\rm cloud} = V_{\rm K}
\left(
\begin{array}{c}
-\sin\phi_{\rm _{B}} \cos\phi_{\rm _{C}} - \cos\phi_{\rm _{B}}\sin\phi_{\rm _{C}}\cos\theta_{\rm _{C}}\\
-\sin\phi_{\rm _{B}} \sin\phi_{\rm _{C}} + \cos\phi_{\rm _{B}}\cos\phi_{\rm _{C}}\cos\theta_{\rm _{C}}\\
\cos\phi_{\rm _{B}}\sin\theta_{\rm _{C}}
\end{array}
\right),
\end{equation}
where $V_{\rm K} = cr_{_{\rm B}}^{-1/2}$. The half opening angle of the BLR is $\Theta_{\rm _{BLR}}$, 
the inner and outer radii of the BLR are $r_{\rm _{B,i}}$ and $r_{\rm _{B,o}}$, respectively. Then we have 
$r_{\rm _{B}} \in [r_{\rm _{B,i}},r_{\rm _{B,o}}], \theta_{\rm _{C}} \in [0,\Theta_{\rm _{BLR}}]$.
The distribution of clouds can be modelled by power law as well. 
The number density of the clouds is
%
%\begin{equation}
$n_{\rm _{B}}(r_{\rm _{B}}, \theta_{\rm _{C}},\phi_{\rm _{C}},\phi_{\rm _{B}}) 
= n_{\rm _{B0}} \left(r_{\rm _{B}}/r_{\rm _{B,i}}\right)^{-\beta}$,
%\end{equation}
%
where $n_{\rm _{B0}}$ is the number density at $r_{\rm _{B,i}}$.

A single scattering process is illustrated by Fig. 1{\it c}. 
Expressions of the scattering angle $\Theta$ and rotation angle $\chi$ are derived in Appendix.
Assuming the ionizing source is isotropic and line intensity of one cloud at $B$ is linearly proportional 
to the intensity of local ionizing fluxes, we have the line intensity at $B$ is $i_{\rm _{B}} = kr_{\rm _{B}}^{-2}$,
where $k$ is a constant. We further assume that all clouds emit unpolarized H$\beta$ photons isotropically and 
neglect multiple scatterings of optically thin regions. We thus have
the intensity at $P$ is simply given by $i_{\rm _{P}} = i_{\rm _{B}} \calS / 4\pi r_{\rm _{BP}}^2$ where $\calS$ 
is the surface area of the cloud. With incident photons with the Stokes parameters of $(i_{\rm _{P}},0,0,0)$,
we have the Stokes parameters in the $(\vec{n}_{\bot}-\vec{n}_{\parallel})$ frame \citep{chandrasekhar1960radiative}.
\begin{eqnarray}
&\displaystyle{\frac{3\sigma}{8\pi R^2}}&\!\!\!\!\!\!\!
\left(\!\!
\begin{array}{cccc}
\frac{1}{2}(1+\cos^2\Theta) & \frac{1}{2}(1-\cos^2\Theta) & 0 & 0\\
\frac{1}{2}(1-\cos^2\Theta) & \frac{1}{2}(1+\cos^2\Theta) & 0 & 0\\
0 & 0 & \cos\Theta & 0 \\
0 & 0 & 0 & \cos\Theta
\end{array}
\!\!\right)
\left(\!\!
\begin{array}{c}
i_{\rm _{P}} \\ 0 \\ 0 \\ 0
\end{array}
\!\!\right)  \nonumber \\
&=&
\calA_0\left(
\begin{array}{c}
1+\cos^2\Theta \\ 1-\cos^2\Theta \\ 0 \\ 0
\end{array}
\right),
\end{eqnarray}
where $\calA_0=3\sigma i_{\rm _{P}}/16\pi R^2$, $R$ is the distance between observer and AGN. Converting 
the Stokes parameter to the fixed coordinate $\vec{n}_{z'}-\vec{n}_{x'}$ system, we have
\begin{eqnarray}
\left(\!\!
\begin{array}{c}
i \\ q \\ u \\v
\end{array}\!\!\right)
&=&\calA_0
\left(\!\!
\begin{array}{cccc}
1 & 0 & 0 & 0\\
0 & \cos 2\chi & \sin 2\chi & 0\\
0 & -\sin 2\chi & \cos 2\chi & 0 \\
0 & 0 & 0 & \cos\Theta
\end{array}\!\!\right)\!
\left(\!\!
\begin{array}{c}
1+\cos^2\Theta \\ 1-\cos^2\Theta \\ 0 \\ 0
\end{array}\!\!\right)
\nonumber \\
&=&\calA_0
\left(\!\!
\begin{array}{c}
1+\cos^2\Theta \\ (1-\cos^2\Theta)\cos 2\chi \\ -(1-\cos^2\Theta)\sin 2\chi \\ 0
\end{array}\!\!\right).
\end{eqnarray}
The velocity of the cloud $\vec{v}_{\rm cloud}$ projected to the direction of incident light $\vec{n}_{\rm _{BP}}$ is
\begin{equation}
\frac{V_{\parallel}}{c} = \frac{1}{r_{\rm _{B}}^{1/2}} \frac{r_{\rm _{P}}(q_1\cos\phi_{\rm _{B}} + q_2\sin\phi_{\rm _{B}})}{[r_{\rm _{P}}^2 + r_{\rm _{B}}^2 + 2r_{\rm _{B}} r_{\rm _{P}}(q_2 \cos\phi_{\rm _{B}} - q_1 \sin \phi_{\rm _{B}})]^{1/2}},
\end{equation}
where $q_1 = \cos\theta_{\rm _{P}} \sin\theta_{\rm _{C}} + \sin\theta_{\rm _{P}}\cos\theta_{\rm _{C}}\sin(\phi_{\rm _{P}} - \phi_{\rm _{C}})$,
$q_2 = -\sin\theta_{\rm _{P}}\cos(\phi_{\rm _{P}} - \phi_{\rm _{C}})$. 
If the intrinsic wavelength of the line is $\lambda_0$, the wavelength after scattering is 
$\lambda' = \lambda_0(1-V_{\parallel}/c)$ due to Doppler shifts.
Integrating over all the clouds and electrons, we have total polarized spectrum,
\begin{equation}
\left(\!\!
\begin{array}{c}
I_{\lambda} \\ Q_{\lambda} \\ U_{\lambda} \\ V_{\lambda}
\end{array}
\!\!\right)
= \int_{V_{\rm _{P}}} dV_{\rm _{P}}n_{\rm _{P}}  \int_{V_{\rm _{BLR}}}\!\!\!\! dV_{\rm _{BLR}} n_{\rm _{B}} 
  \int_{0}^{2\pi} {\cal L}(\lambda,\lambda^{\prime}) d\phi_{\rm _{B}}
\left(
\begin{array}{c}
i \\ q \\ u \\ v
\end{array}
\right)
\end{equation}
where the intrinsic profile of H$\beta$ line is assumed to be a Lorentzian function of
${\cal L}(\lambda,\lambda^{\prime})\propto \Gamma/[(\lambda-\lambda_0)^2+\Gamma^2]$, $\Gamma$ 
is the intrinsic width much smaller than the broadening due to rotation of clouds 
(the Lorentzian profiles are a very good approximation in the present context).

Similarly, we can calculate the spectrum of non-scattered photons. The velocity of the cloud $\vec{v}_{\rm cloud}$ projected to the 
direction of observer $\vec{n}_{\rm obs}$ is
\begin{equation}
\frac{\tilde{V}_{\parallel}}{c} = \frac{1}{r_{\rm _{B}}^{1/2}} \left( \tilde{q}_1\cos\phi_{\rm _{B}} + \tilde{q}_2\sin\phi_{\rm _{B}} \right),
\end{equation}
where $\tilde{q}_1 = \sin\theta_{\rm _{C}} \cos i + \cos\theta_{\rm _{C}}\cos\phi_{\rm _{C}} \sin i$, $ \tilde{q}_2 = -\sin\phi_{\rm _{C}}\sin i$. 
The observed wavelength is $\lambda'' = \lambda_0(1-\tilde{V}_{\parallel}/c)$. 
Integrating over all the clouds, we have
\begin{equation}
F_{\lambda} =
\int_{V_{\rm _{BLR}}} dV_{\rm _{BLR}} n_{\rm _{B}} \int_{0}^{2\pi} {\cal L}(\lambda,\lambda^{\prime\prime})
 d\phi_{\rm _{B}} \frac{i_{\rm _{B}}S}{4\pi R^2},
\end{equation}
and the expression for polarization degree and position angle,
\begin{equation}
P_{\lambda} = \frac{\sqrt{Q_{\lambda}^2 + U_{\lambda}^2}}{I_{\lambda} + F_{\lambda}}, \quad
\theta_{\lambda} = \frac{1}{2} \arccos\left( \frac{Q_{\lambda}}{\sqrt{Q_{\lambda}^2+U_{\lambda}^2}} \right).
\end{equation}
If $U_{\lambda} > 0$, $\theta_{\lambda} \in (0,\pi/2)$. If $U_{\lambda} < 0$, 
$\theta_{\lambda} \in (-\pi/2,0)$. Position angles represent the angle 
between the direction of maximum intensity and  $n_{z'}$.

\begin{figure}
\centering
\includegraphics[scale=0.17, trim = 95 85 100 65]{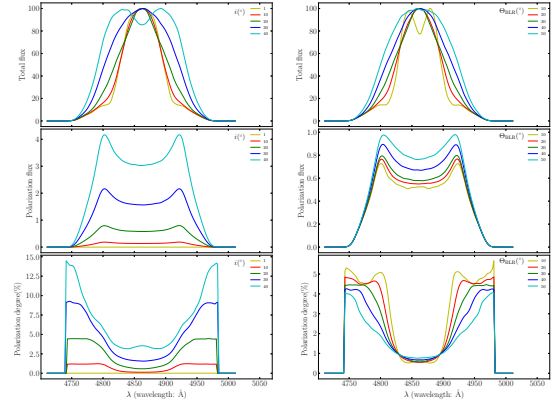}
\caption{
Total, polarised spectra and polarisation degrees of a type I AGN with an equatorial electron 
scattering region. The left column are spectra for different inclinations whereas the right 
for different BLR opening angles. The tops of the total spectra become flatter and 
polarisation decreases with increases $\Theta_{\rm BLR}$ (tends to $90^{\circ}$). }
\end{figure}

Table 1 summarises all the parameters of the present model. We calculate a series of profiles for different
values of parameters and find that the profiles are only sensitive to $\Theta_{_{\rm BLR}}$ and $i$.
Fig. 2 shows typical spectra of a type I AGN with an equatorial 
electron scattering region. The parameters of the model are 
$r_{\rm _{P,i}} = 10^4$, $r_{\rm _{P,o}} = 2\times10^4$, $\Theta_{\rm _{P}} = 30^{\circ}$, 
$r_{\rm _{B,i}} = 2\times10^3$, $r_{\rm _{B,o}} = 6\times10^3$,
$\alpha = 1$, $\beta = 1$, $\tau_{\rm es}=\sigma n_{\rm _{P0}}(r_{\rm _{P,o}}-r_{\rm _{P,i}}) = 1$,
$i =(1^{\circ},10^{\circ},20^{\circ},30^{\circ}, 40^{\circ})$ and 
$\Theta_{_{\rm BLR}}  =(10^{\circ},20^{\circ},30^{\circ},40^{\circ},50^{\circ})$. 
As shown by the left panel of Fig. 2, the total spectra get broader as $i$ increases and show double peaked-profiles 
when $i$ exceeds a critical inclination. By contrast, the width and the profile of the polarised spectra are not
sensitive to $i$ (can be found from normalized spectra). This interesting property results from the fact that 
the polarised spectra are equivalent to the ones seen by observers at edge-on orientations. Generally, the line 
centres have lowest polarisation degrees. However, polarization degrees become smaller with $i$.

Total spectra show strong dependence on $\Theta_{\rm BLR}$. As shown in the right panel of Fig. 2, 
large-$\Theta_{\rm BLR}$ BLRs show broader width and get narrower with decreases of $\Theta_{\rm BLR}$ 
until double peaked-profiles.
However, $\Theta_{_{\rm BLR}} $ does not change the polarised profiles too much. Large $\Theta_{_{\rm BLR}}$ 
indicates that the system tends to be more spherically symmetric and to decrease the polarization degree.
Comparing with the total spectra, the width of the polarised spectrum is insensitive to $\Theta_{_{\rm BLR}}$ 
and $i$. This property allows us to infer $\fBLR$ from the polarized spectrum and improve 
the accuracy of BH mass measurement as shown in \S3.

\section{The virial factor}
With the BLR model, we have the emissivity-averaged time lag of broad emission line as
\begin{equation}
\Delta t = \frac{\int dV_{\rm _{BLR}} \Delta r i_{\rm _{B}} n_{\rm _{B}}}{\int dV_{\rm _{BLR}} i_{\rm _{B}} n_{\rm _{B}}} \left(\frac{R_{\rm g}}{c}\right)
= \frac{1-\beta}{2-\beta} \frac{1-q_r^{2-\beta}}{1-q_r^{1-\beta}} \left(\frac{r_{_{\rm B,i}}R_{\rm g}}{c}\right),
\end{equation}
where $\Delta r = r_{\rm _{B}} - \vec{r}_{\rm _{B}}\cdot\vec{n}_{\rm obs}$ and the corresponding virial factor from Eq. (1) can be written as 
\begin{equation}
\fBLR^{-1} =
\left(\frac{1-\beta}{2-\beta}\right) \left(\frac{1-q_r^{1-\beta}}{1-q_r^{2-\beta}}\right) 
\left(\frac{V_{\rm _{FWHM}}}{c}\right)^2r_{_{\rm B,i}},
\end{equation}
where $q_r=r_{_{\rm B,o}}/r_{_{\rm B,i}}$, $\VFWHM$ is FWHM of profiles either from the total 
or polarised spectra. We generate profiles according to parameters listed in Table 1 and measure 
$\VFWHM$ to show dependences of $\fBLR$ on each parameter.

We did Monte-Carlo simulations for all the parameters listed in Table 1 and then get the 
$\ln\fBLR-\ln X_i$ relations, where $X_i$ is any one of the parameters. The $\fBLR$-dependence 
on $X_i$ can then be obtained by 
$\delta \ln\fBLR=\left(\partial \ln \fBLR/\partial \ln X_i\right)\delta \ln X_i$, 
where $\partial \ln \fBLR/\partial \ln X_i$ is the slope of the $\ln\fBLR-\ln X_i$ relations,
$\delta \ln X_i$ is its range. The slope is estimated from the line regression of $\ln \fBLR-\ln X_i$ 
relations from the Monte-Carlo simulations. Dependence listed
in Table 1 shows that only $i$ and $\Theta_{\rm BLR}$ are the major drivers in the total
spectra, but $\fBLR$ is insensitive to all the parameters (only slightly relies on $\Theta_{\rm BLR}$). 
We estimate the entire uncertainties of $\fBLR$ due to all parameters as
$\Delta \log \fBLR=\left[\sum_{i=1}^9 (\partial \ln \fBLR/\partial \ln X_i)^2\left(\Delta \log X_i\right)^2\right]^{1/2}$.
We have $\Delta \log \fBLR=(0.74,0.04)$ for total and polarised spectra, respectively.

We plot the $\log \fBLR-$distributions and contour maps
versus $\Theta_{\rm BLR}$ and $i$ in Fig. 3. It shows that the distribution from total spectra 
is much broader than that from polarised spectra. The $68\%$ confidence interval of the former is 
$\log\fBLR\in[-0.41,0.08]$ agreeing with values from detailed MCMC modelling \citep{pancoast2014modelling2},
but $\log\fBLR\in[-0.65,-0.62]$ from the polarised spectra. For a typical RM campaign, we have the 
uncertainties of BH mass $\delta\bhm/\bhm\approx (0.8, 0.3)$ from total and polarised spectra, 
respectively. Obviously, spectropolarimetry provides much better $\fBLR$ for BH mass from RM campaign.
However, the polarised spectra as the prerequisites of applications should be identified as 
originating from the equatorial scatters. 
Actually, this can be done by checking if the position angles are parallel to radio axis in AGNs.

We apply the current $\fBLR$ to the radio-loud narrow line Seyfert 1 galaxy PKS 2004$-$447 with
$V_{\rm FWHM}({\rm H\alpha})=1500\,\kms$($z=0.240$). The black hole mass estimated by single total 
spectra is about $5 \times 10^6 M_{\odot}$ \citep{oshlack2001very},
which is much smaller than the critical mass ($\bhm\sim 10^8\sunm$) invariably associated with classical 
radio load(RL) AGNs\citep{laor2000black}. Fortunately, the polarised spectra of VLT observations show its 
$\rm H\alpha$ FWHM of $(280 \pm 50)$\AA\,\citep{baldi2016radio}.
Since the 5100\AA\, luminosity is $L_{5100} = 1.25 \times 10^{44} \rm erg \, s^{-1}$ \citep{gallo2006spectral},
$R-L$ relation indicates that the average time lag between emission line and continuum is about $10^{1.6\pm 0.14}$ 
days \citep{bentz2013low} for sub-Eddington AGNs \citep{du2016supermassive}.
Taking $\log \fBLR = -0.63$ for the polarised spectra of H$\alpha$ line, we have $\bhm= 10^{8.45 \pm 0.2}\sunm$.
Employing the standard accretion disk model, we have the dimensionless accretion rates of 
$\dot{\mathscr{M}}=\dot{M}c^2/L_{\rm Edd}=20.1\left(L_{44}/\cos i\right)^{3/2}m_7^{-2}\approx 0.05$, where
$\dot{M}$ is the accretion rates, $L_{\rm Edd}=1.4\times 10^{38}\left(\bhm/\sunm\right)\ergs$, 
$L_{44}=L_{5100}/10^{44}\ergs$, $\cos i=0.75$ (inclination) and $m_7=\bhm/10^7\sunm$ \citep{du2014supermassive}. 
Such a low accretion rate agrees with the radio-loudness and accretion rate 
relation \citep{sikora2007radio}. % \citep{ho2002relationship, sikora2007radio}.

Finally, we would like to point out the temporal properties of polarised spectra. The equatorial distributions 
of scatters lead to delays of polarised photons with different frequencies relative to the BLR, and such a delay 
may need to be considered for polarised spectra at different epochs.
Such a kind of polarisation campaigns will provide a new way of accurately  measuring 
the black hole mass in type 1 AGNs.

\begin{deluxetable}{lllcc}
\tablecolumns{5}
\tablewidth{1pt}
\setlength{\tabcolsep}{3pt}
\tablewidth{0pc}
\tablecaption{The dependence of $\fBLR$ on model parameters
~~~~~~~~~~~~~~~~~~~~~~~~~~~~~~~~~~~~~~~~~~~~~~~~~~~~~~
~~~~~~~~~~~~~~~~~~~~~~~~~~~~~~}
\tabletypesize{\footnotesize}
\tablehead{
\colhead{Parameter} &
\colhead{Range}     &
\colhead{Meaning}   &
\multicolumn2c{$\delta\ln f_{_{\rm BLR}}$}\\ \cline{4-5}
\colhead{}          &
\colhead{}          &
\colhead{}          &
\colhead{total}     &
\colhead{polarized}
}
\startdata
$r_{\rm _{P,i}}(10^4 R_{\rm g})$ & $[1,5]$ & SR inner radius & $0.01$ & $0$\\
    $r_{\rm _{P,o}}(10^4 R_{\rm g})$ & $[2,10]$ & SR outer radius & $0$ & $0$\\
    $\Theta_{\rm _{P}}(^\circ)$  & $[20,50]$ & SR opening angle & $-0.04$ & $0$\\
    $\alpha$ & $[0,1.5]$ & index of DF of electrons & $0.01$ & $0$\\
    $r_{\rm _{B,i}}(10^3 R_{\rm g})$ & $[1,5]$ & inner radius of BLR & $-0.01$ & $0.01$\\
    $r_{\rm _{B,o}}(10^3 R_{\rm g})$ & $[2,10]$ & outer radius of BLR & $0.01$ & $-0.01$\\
    $\Theta_{\rm _{BLR}}(^\circ)$ & $[20,50]$ & opening angle of BLR & $-0.38$ & $0.03$\\
    $\beta$ & $[0,1.5]$ & index of DF of clouds & $0$ & $-0.02$ \\
    $i(^\circ)$ & $[0,45]$ & inclination angle & $-0.64$ & $0$ \\
\enddata
\vglue -0.65cm
\tablecomments{SR: scattering region, DF: distribution function.
$\delta \ln\fBLR$ describes the dependence on parameters of the broad-line
regions. See details for its definition in the main text. It shows that $\fBLR$ is sensitive to
$\Theta_{\rm BLR}$ and $i$ for the total spectra, but it is almost a constant for polarised spectra.
}
\end{deluxetable}

\begin{figure}
\centering
\includegraphics[scale=0.38,trim = 50 30 80 40]{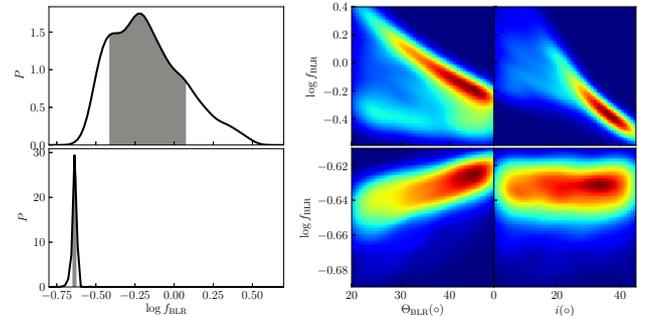}
\caption{
The upper three panels display the distribution of $\log \fBLR$ obtained from total spectra while the lower three 
polarized spectra. In each row, the first panel is the probability density function of $\log \fBLR$ and the shadow 
area marks out the $68\%$ confidence interval, the second and third indicate correlations of 
$\log \fBLR$ with model parameters of $\Theta_{\rm _{BLR}}$ and $i$, respectively. $\fBLR$ is sensitive to both
$\Theta_{\rm BLR}$ and $i$ for the total spectra whereas $\fBLR$ only very weakly depends on $\Theta_{\rm BLR}$. }
\end{figure}

\section{Conclusion and Discussion}
In this Letter, we show that the factor $f_{\rm BLR}$ has a wide range for total spectra of the 
virialized the broad-line regions. We investigate the polarised spectra of the BLR arisen from the 
equatorial scatters for $\fBLR$ in determination of black hole mass in type I AGNs. It is found
that $\log\fBLR\in[-0.65,-0.62]$ for polarised spectra. This arises from the fact that
the electrons on the equatorial plane scatter the broad-line photons to observers, assembling to 
view the BLR as edge-on orientation. The polarised spectra provide a way of accurately measuring BH 
mass from single epoch polarised spectra, which is much better than that from single epoch total spectra.
For an individual application, equatorial scatters must be checked for the validity of the polarised spectra.

We note the work of \citet{afanasiev2015polarization}, who employed the angles of polarisation arisen by 
scatters on the inner edge of the dusty torus in order to alleviate dependence of black hole mass on inclinations.
This is different from what we suggest in this paper.
We would like to point out the major assumptions used in this paper
that scattering region is equatorial, but static. The geometry of scatters 
is supported by observations but the dynamics could be more complicated. We also neglect scatters of dust particles.
This simple model shows the potential functions of polarised spectra in measuring black hole mass. Fitting the
polarised spectra leads to more accurate BH mass, but we will conduct it in a forthcoming paper.

\section*{Acknowledgements}
The authors thanks the referee for a useful report. C. Tao is thanked for useful discussions.
We acknowledge the support of the staff of the Lijiang 2.4 m telescope. Funding for the telescope has been 
provided by CAS and the People's Government of Yunnan Province. This research is supported by National Key 
Program for Science and Technology Research and Development (grant 2016YFA0400701),
NSFC grants through NSFC-11503026, -11173023, and -11233003, and a NSFC-CAS joint key grant U1431228, by 
the CAS Key Research Program through KJZD-EW-M06, and by Key Research Program of Frontier Sciences, CAS, 
grant No. QYZDJ-SSW-SLH007.

%%%%%%%%%%%%%%%%%%%%%%%%%%%%%%%%%%%%%%%%%%%%%%%%%%

%%%%%%%%%%%%%%%%%%%% REFERENCES %%%%%%%%%%%%%%%%%%

%%%%%%%%%%%%%%%%%%%%%%%%%%%%%%%%%%%%%%%%%%%%%%%%%%f

%%%%%%%%%%%%%%%%% APPENDICES %%%%%%%%%%%%%%%%%%%%%

\appendix
\section*{Appendix}

For type I AGN, the polarised spectra observed by a remote observer are mostly caused by electron scattering 
of the equatorial regions. Suppose the coordinate of the electron in spherical system is $(r_{\rm _{P}}$, 
$\theta_{\rm _{P}}$, $\phi_{\rm _{P}})$, we have
\begin{equation}
\begin{cases}
x_{\rm _{P}} = r_{\rm _{P}}\sin\theta_{\rm _{P}}\cos\phi_{\rm _{P}}, \\
y_{\rm _{P}} = r_{\rm _{P}}\sin\theta_{\rm _{P}}\sin\phi_{\rm _{P}}, \\
z_{\rm _{P}} = r_{\rm _{P}}\cos\theta_{\rm _{P}}.
\end{cases}
\end{equation}
As for a cloud in BLR at $(r_{\rm _{B}}\cos\phi_{\rm _{B}}, r_{\rm _{B}}\sin\phi_{\rm _{B}},0)$, we
use the rotation matrix 
\begin{equation}
\mathbf{R} =
\left(
\begin{array}{ccc}
\cos\phi_{\rm _{C}} & -\sin\phi_{\rm _{C}}\cos\theta_{\rm _{C}} & \sin\phi_{\rm _{C}}\sin\theta_{\rm _{C}} \\
\sin\phi_{\rm _{C}} &  \cos\phi_{\rm _{C}}\cos\theta_{\rm _{C}} & -\cos\phi_{\rm _{C}}\sin\theta_{\rm _{C}} \\
0 & \sin\theta_{\rm _{C}} & \cos\theta_{\rm _{C}}
\end{array}
\right),
\end{equation}
for the position of the cloud 
\begin{equation}
\begin{cases}
x_{\rm _{B}} = r_{\rm _{B}} (\cos\phi_{\rm _{B}}\cos\phi_{\rm _{C}} - \sin\phi_{\rm _{B}}\sin\phi_{\rm _{C}}\cos\theta_{\rm _{C}}),\\
y_{\rm _{B}} = r_{\rm _{B}} (\cos\phi_{\rm _{B}}\sin\phi_{\rm _{C}} - \sin\phi_{\rm _{B}}\cos\phi_{\rm _{C}}\cos\theta_{\rm _{C}}),\\
z_{\rm _{B}} = r_{\rm _{B}}\sin\theta_{\rm _{C}}\sin\phi_{\rm _{B}},
\end{cases}
\end{equation}
in the $O-XYZ$ frame.

%For a single scattering, suppose that the coordinate for 
One cloud is at 
$(x_{\rm _{B}},y_{\rm _{B}},z_{\rm _{B}})$ and one scattering electron is at 
$(x_{\rm _{P}},y_{\rm _{P}},z_{\rm _{P}})$, we have the direction vector of incident light
\begin{equation}
\vec{n}_{\rm _{BP}} = \frac{1}{r_{\rm _{BP}}} \left[(x_{\rm _{P}}-x_{\rm _{B}})\vec{i} +  (y_{\rm _{P}}-y_{\rm _{B}})\vec{j} + (z_{\rm _{P}}-z_{\rm _{B}})\vec{k}\right],
\end{equation}
where $r_{\rm _{BP}} = \sqrt{(x_{\rm _{P}}-x_{\rm _{B}})^2 + (y_{\rm _{P}}-y_{\rm _{B}})^2 + (z_{\rm _{P}}-z_{\rm _{B}})^2}$.
The direction of the observer at infinity is taken to be $\vec{n}_{\rm obs} = (0,\sin i, \cos i)$. 
The scattering angle is given by
\begin{equation}
\cos\Theta = \vec{n}_{\rm obs} \cdot \vec{n}_{\rm _{BP}} = \frac{1}{r_{\rm _{BP}}} \left[ (y_{\rm _{P}} - y_{\rm _{B}})\sin i + (z_{\rm _{P}} - z_{\rm _{B}})\cos i\right].
\end{equation}
The unit vector perpendicular to the scattering plane is
\begin{equation}
\vec{n}_{\bot} = \frac{\vec{n}_{\rm _{BP}} \times \vec{n}_{\rm obs}}{|\vec{n}_{\rm _{BP}} \times \vec{n}_{\rm obs}|} .
\end{equation}
We take the fixed coordinate system at celestial sphere of the observer to be $\vec{n}_{z'}-\vec{n}_{x'}$, $\vec{n}_{z'} = (0,-\cos i, \sin i),\vec{n}_{x'} = (1,0, 0)$. 
The angle between $\vec{n}_{\bot}$ and $\vec{n}_{z'}$ satisfies that
\begin{equation}
\cos \chi =  \frac{x_{_{\rm P}}-x_{_{\rm B}}}{\left\{[(z_{\rm _{P}}-z_{\rm _{B}})\sin i - (y_{\rm _{P}}-y_{\rm _{B}})\cos i]^2 + (x_{\mathrm{P}}-x_{\mathrm{B}})^2\right\}^{1/2}}. 
\end{equation}

%%%%%%%%%%%%%%%%%%%%%%%%%%%%%%%%%%%%%%%%%%%%%%%%%%

% Don't change these lines
\bsp	% typesetting comment
\label{lastpage}
\end{document}